\documentclass{PoS}

\title{Magnetized QCD phase diagram: critical end points for the strange quark phase transition driven by external magnetic fields}

\ShortTitle{CEP for the strange quark phase transition driven by external magnetic fields}

\author{\speaker{Pedro Costa}, M\'arcio Ferreira and Constan\c ca Provid\^encia\\
        CFisUC, Department of Physics, University of Coimbra, P-3004 - 516 Coimbra, Portugal\\
        E-mail: \email{pcosta@uc.pt}, \email{mferreira@teor.ﬁs.uc.pt}, \email{cp@fis.uc.pt}}

\abstract{
In this work we examine possible effects of an external magnetic field in the 
strongly interacting matter phase diagram. The study is performed using the 
Polyakov-Nambu-Jona-Lasinio model. Possible consequences of the inverse 
magnetic catalysis effect on the phase diagram at both finite chemical 
potential and temperature are analyzed. 
We devote special emphasis on how the location of the multiple critical end 
points (CEPs) change in a magnetized medium: the presence of an external 
magnetic field induces several CEPs in the strange sector, which arise due to 
the multiple phase transitions that the strange quark undergoes.
We also study the deconfinement transition which turns out to be less sensitive 
to the external magnetic field when compared to the quark phase transitions. 
The crossover nature of the deconfinement is preserved over the whole 
phase diagram.
}

\FullConference{XVII International Conference on Hadron Spectroscopy and Structure - Hadron2017\\
		25-29 September, 2017\\
		University of Salamanca, Salamanca, Spain}

\begin{document}

\begin{figure}[!htbp]
	\centering
	\includegraphics[width=0.7\linewidth]{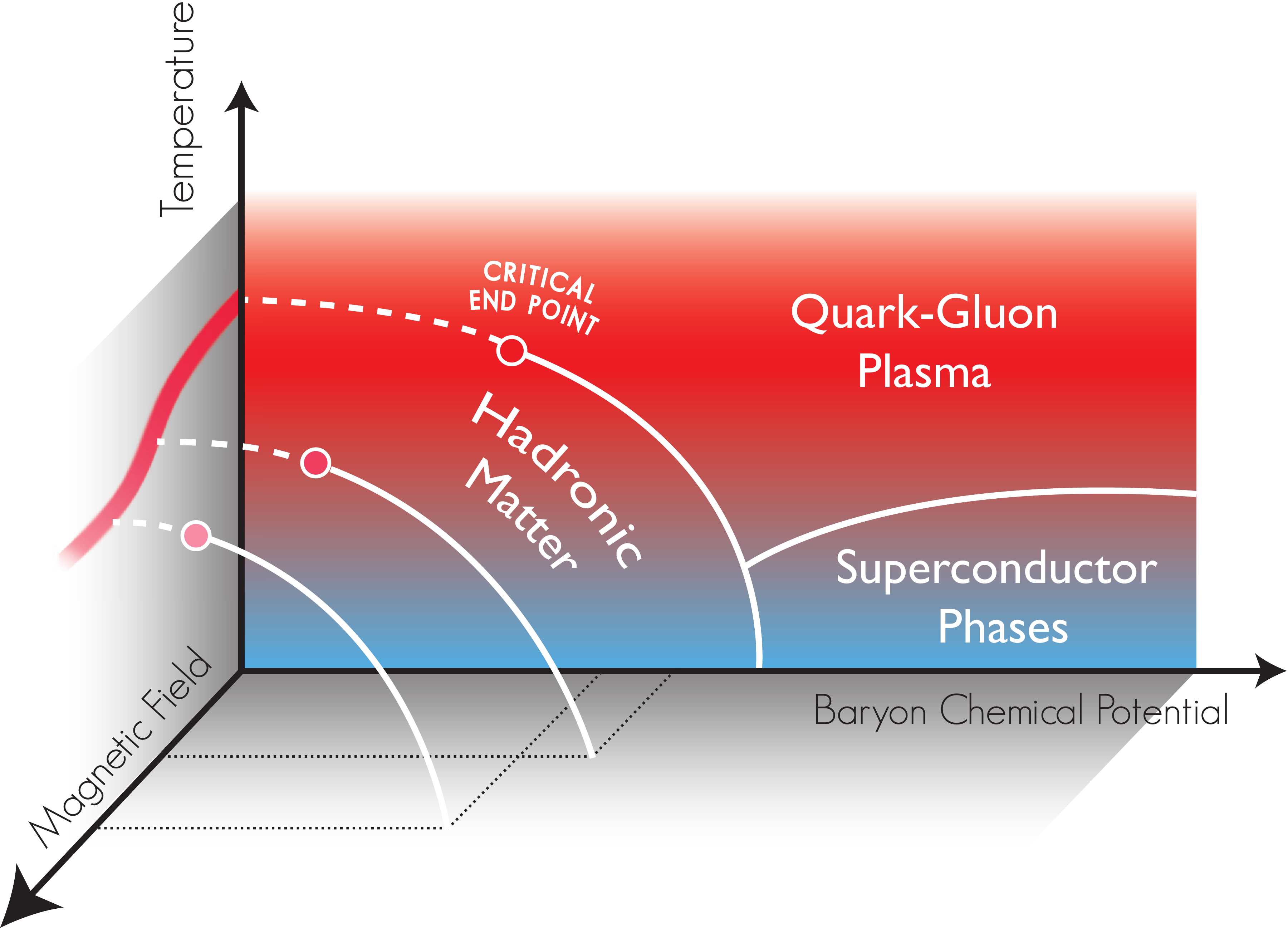}
	\caption{Schematic structure of the QCD matter in the presence of an external
	magnetic field.} 
	\label{fig:Fig1}
\end{figure}

The QCD phase diagram and the respective chiral critical end point (CEP), 
belong to a set of quantum field theoretical phenomena that are affected by 
the presence of external magnetic fields (see Fig. \ref{fig:Fig1}) 
\cite{Andersen:2014xxa}. 
A great attention has recently been given to this subject 
\cite{Ferreira:2013tba,Ferreira:2014kpa,Ferreira:2013oda,Ferreira:2014exa} 
due to its relevance for heavy ion collisions (HIC) measurements 
\cite{Skokov:2009qp}, for the physics of compact stars \cite{Menezes:2014aka}, 
and for the understanding of the primordial stages of the universe 
\cite{Enqvist:1993np}.
Having this in mind, different scenarios involving regions of the phase diagram 
in the presence of external magnetic fields were studied using the 
Polyakov--Nambu-Jona-Lasinio (PNJL) model with (2+1)-flavors 
\cite{Costa:2013zca,Costa:2015bza,Costa:2016vbb}. 
In some of these works, it was analyzed how the location of the CEP depends on 
the presence of magnetic fields.
In \cite{Costa:2013zca}, for example, it was shown that large isospin asymmetry 
moves the CEP to smaller temperatures leading, eventually, to its disappearance 
from the phase diagram.
Nevertheless, a first-order phase transition will be restored in the phase 
diagram  if a strong enough magnetic field is present. 

One well known and understood mechanism induced by the presence of an external 
magnetic field is the catalyzing effect on the dynamical chiral symmetry 
breaking, the so-called Magnetic Catalysis (MC) effect \cite{Klevansky:1989vi}. 
Lattice QCD (LQCD) studies at finite temperature  have shown, however, that the 
magnetic field has an interesting behavior in the transition temperature 
region: instead of catalyzing, it weakens the dynamical chiral symmetry 
breaking, the so-called Inverse Magnetic Catalysis (IMC) \cite{Bali:2011qj}.
Several explanations have been proposed to clarify this unexpected effect 
\cite{Andersen:2014xxa}.
Motivated by LQCD calculations reported in \cite{Bruckmann:2013oba}, the IMC 
effect was incorporated successfully for the first time in 
\cite{Ferreira:2013tba}: with the introduction of an indirect weakening of the 
model scalar coupling, $G_s$, with $B$ (via the Polyakov potential), it was 
obtained an extended (2+1)-PNJL model that presented an IMC effect for the 
quarks condensates at finite temperature.
Later in \cite{Ferreira:2014kpa}, we considered the screening effects of strong 
interactions through the scalar coupling ($G_s(eB)$), achieving a qualitative 
agreement with LQCD results. 

\begin{figure}[!htbp]
	\centering
	\includegraphics[width=1\linewidth]{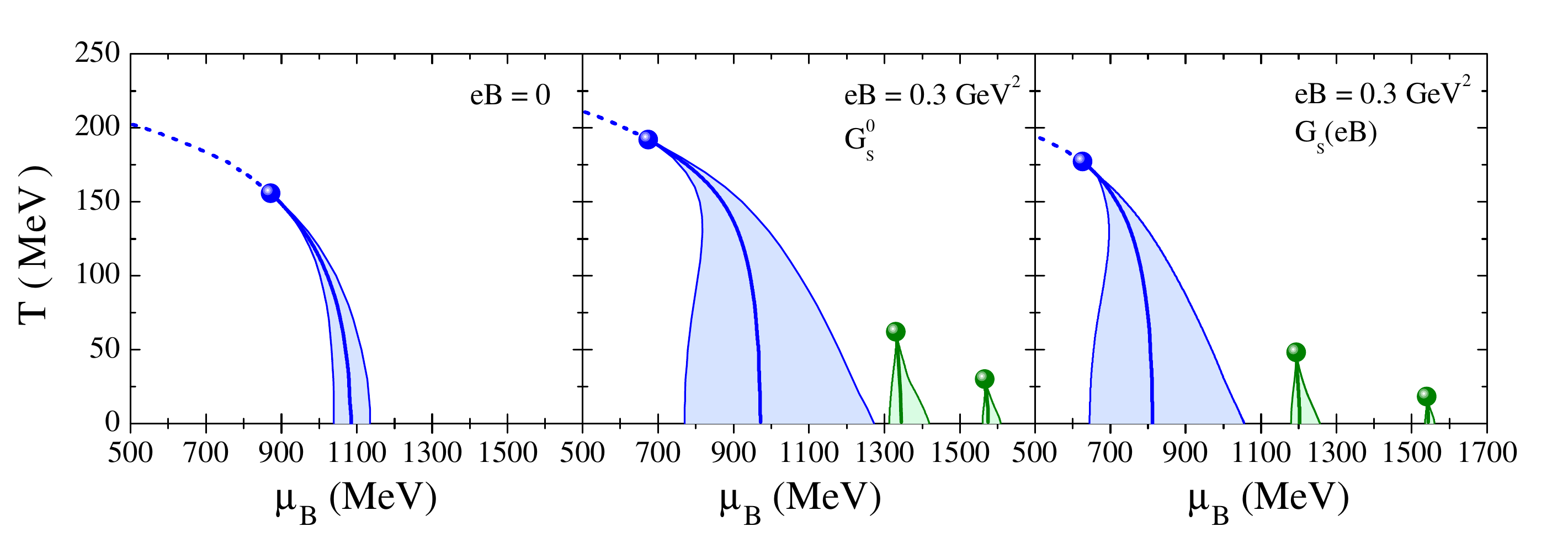}
	\includegraphics[width=1\linewidth]{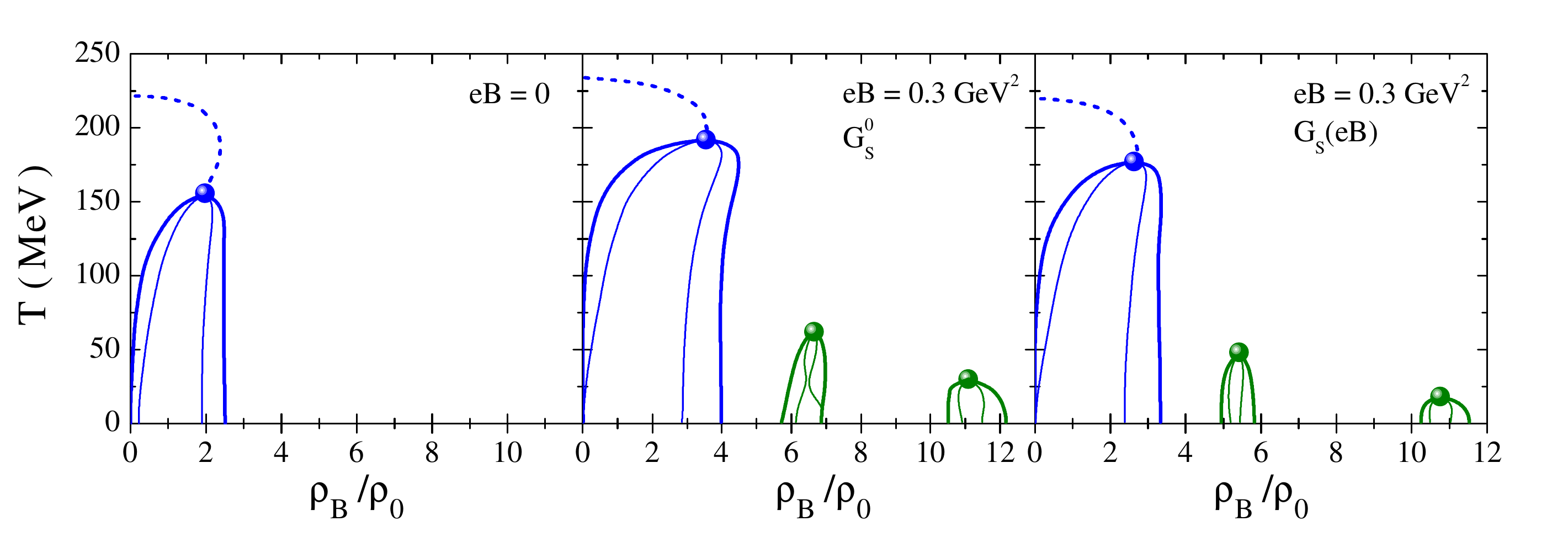}
	\caption{The $T-\mu_B$ (top panels) and $T-\rho_B$ (bottom panels)
	diagrams for: $eB=0$ (left); $eB=0.3$ GeV$^2$ 
	and $G_s^0=const.$, scenario with no IMC mechanism (middle); and $eB=0.3$ 
	GeV$^2$ and $G_s(eB)$, scenario with IMC mechanism (right panel).
	The baryonic density $\rho_B$ is represented in units of saturation
		density, $\rho_0=0.16$ fm$^3$.} 
	\label{fig:Fig2}
\end{figure}

At finite temperature and density/chemical potential, we now study how the IMC 
mechanism, via a magnetic field dependent coupling $G_s(eB)$, affects the 
first order region and the position of the CEP.
In Fig. \ref{fig:Fig2} we present the phase diagram ($T-\mu_B$ plane - upper 
panels; $T-\rho_B$ plane - lower panels) for three different cases: 
$eB=0$ - left panels; 
$eB=0.3$ GeV$^2$ and $G_s^0=const.$ (no IMC effect) - middle 
panels; 
$eB=0.3$ GeV$^2$ and $G_s(eB)$ (with IMC effect) - right panels.
From the upper panels, we conclude that the presence of a magnetic field will: 
a) enlarge the spinodal region for the light sector, being more 
pronounced without an IMC mechanism (middle panel);
b) move the light CEP to lower values of $\mu_B$, being stronger
when the IMC effect is present (right panel); 
c) generate multiple first-order phase transitions for the strange sector with 
the respective appearance of multiple CEPs in this sector 
(for $eB \gtrsim 0.4$ GeV$^2$, only one strange CEP exists).
Instead of a single first-order phase transition connecting the vacuum phase to 
the chirally restored phase, several intermediate first-order phase transitions 
take place that are generated by Landau quantization, induced by the magnetic 
field presence, and a succession of partial restorations of the chiral symmetry. 

Another relevant aspect for both, light and strange, transitions is that for 
stronger magnetic fields the spinodal region is enlarged, being this region 
bigger for $G_s = G_s^0$ \cite{Ferreira:2017}.
The first-order lines are moved to lower baryonic chemical potentials. 
From the bottom panels of Fig. \ref{fig:Fig2}, we also conclude that the upper 
baryonic densities at which the onset of both spinodal  and 
binodal regions occur increase with $B$ for both cases. 
Moreover, the spinodal region for the
strange quark is much smaller than for the light quarks and is located at 
higher values of $\rho_B$.

\begin{figure}[!htbp]
	\centering
	\includegraphics[width=0.5\linewidth]{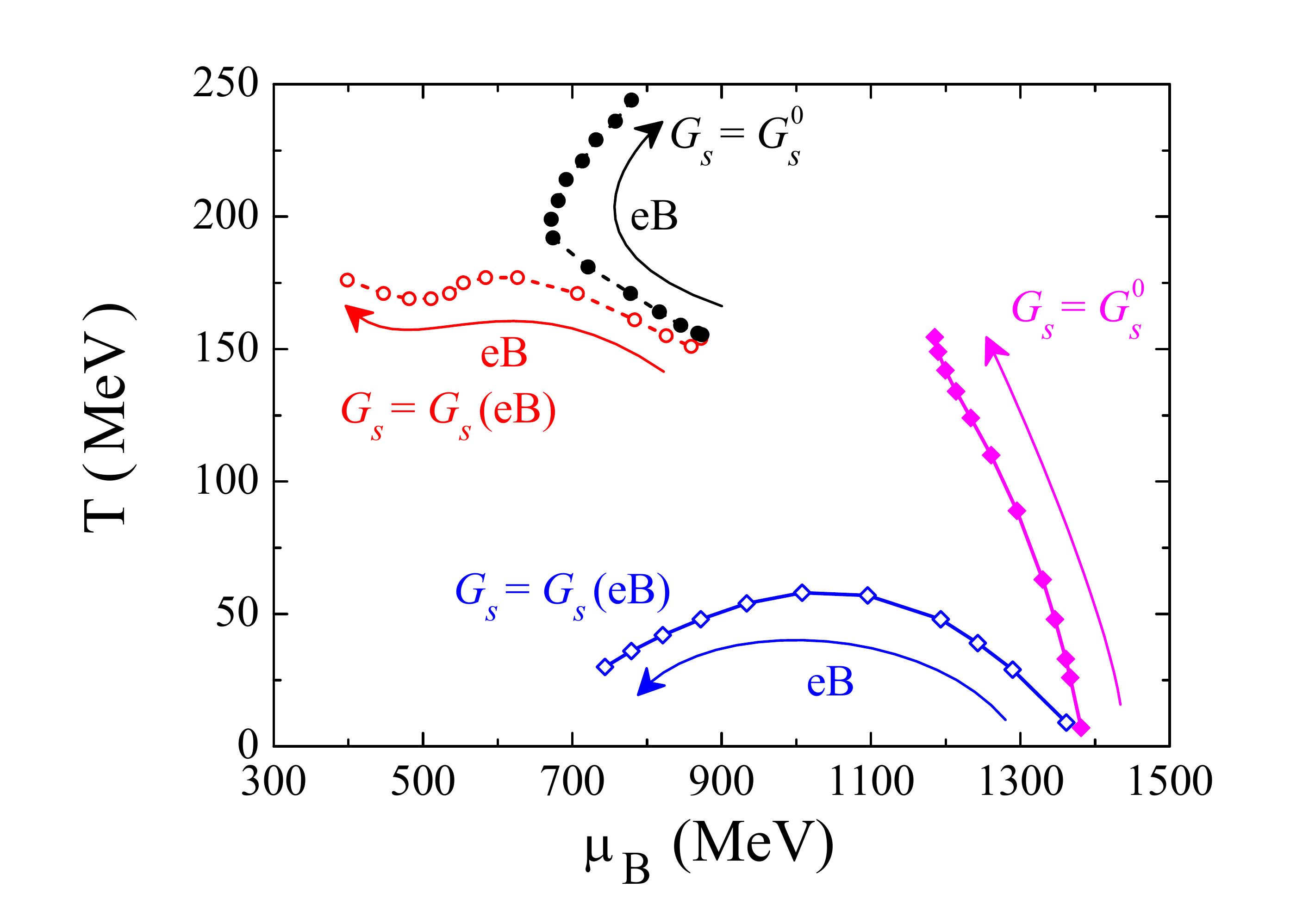}
	\hspace{-0.25cm}\includegraphics[width=0.505\linewidth]{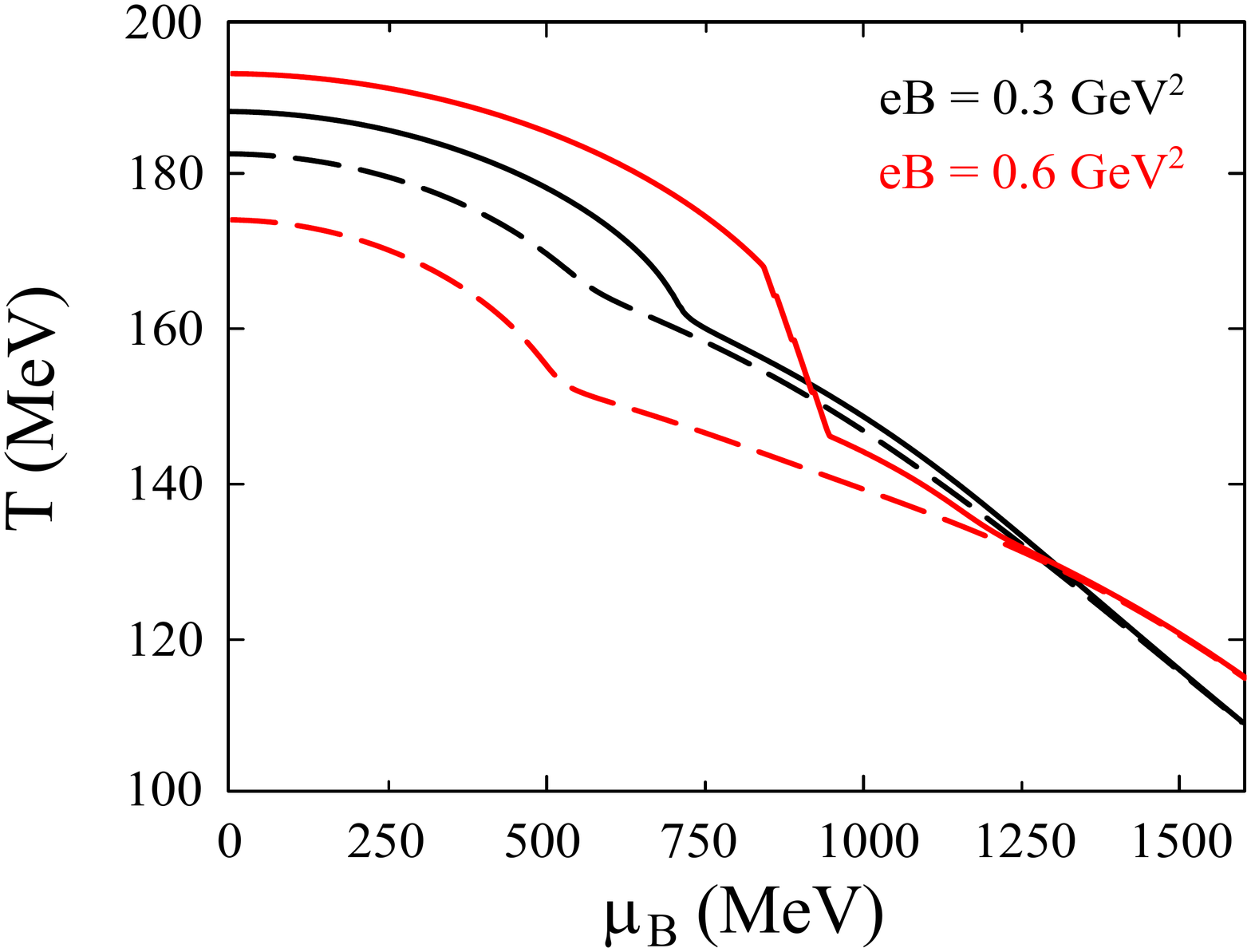}
	\caption{Left panel: CEPs of the light (red and black) and strange (blue and 
	magenta) quarks as a function of $B$ for both scenarios: a constant coupling, 
	$G_s^0$, and 	magnetic dependent coupling, $G_s (eB)$. The magnetic field 
	increases from 0 to 1 GeV$^2$ in the arrows' directions. 
	Right panel: $\Phi(T,\mu_B) = 0.5$ when $G_s=G_s^0$ (full lines) and 	when 
	$G_s(eB)$ (dashed lines) for $eB=0.3$ GeV$^2$ (black lines) and $eB=0.6$ 
	GeV$^2$ (red lines).} 
	\label{fig:Fig3}
\end{figure}

Concerning the CEPs, we present the results in Fig. \ref{fig:Fig3} (left panel). 
We start by comparing the CEP's position for the light sector ($u$ and $d$ 
quarks) with and without the IMC mechanism. 
For magnetic fields lower than $0.3$ GeV$^2$, we have found that the presence 
of an IMC mechanism has a small effect in the CEP position, i.e., the CEPs move 
towards higher values of $T$ and $\mu_B$ in both scenarios (see red and black 
curves). 
For higher magnetic fields, however, the CEP is moved to lower $\mu_B$ with 
increasing magnetic fields for $G_s (eB)$, while the temperature remains almost 
unchanged \cite{Costa:2015bza}. 
Indeed, the $G_s(eB)$ results indicate that, for high enough magnetic fields, 
the CEP goes towards the $\mu_B = 0$ axis, and the crossover transition in this 
axis will eventually turn into a first-order phase transition. 
This behavior is distinct from the one obtained when the IMC effect is absent 
($G_s^0$): 
above a critical magnetic field strength, the CEP location is shifted to higher 
values of $T$ and $\mu_B$ with increasing magnetic field \cite{Costa:2013zca}. 

Let us now focus our attention on the CEP of the strange sector. As we already 
saw, the presence of a magnetic field induces multiple CEPs. 
For both scenarios, we focus only on the CEP appearing at lower 
$\mu_B\,\,\,(\rho_B)$ in Fig. \ref{fig:Fig2} that remains up to 
$eB\sim 1$ GeV$^2$ (the CEP at higher $\mu_B$ disappears from the phase diagram 
at $eB \sim 0.4$ GeV$^2$; similarly to the CEP for light sector 
\cite{Costa:2013zca}). 
The CEP's position shows a different behavior depending on the presence of an 
IMC mechanism: while at lower values of $B$ it moves towards lower $\mu_B$ 
in both scenarios, at high magnetic fields the $T^{CEP}$ increases monotonously 
with the intensity of the magnetic field for a constant coupling $G_s^0$, but 
$T^{CEP}$ is a decreasing function when we have $G_s(eB)$. 

With increasing $B$, the position of the CEP in the scenario with $G_s (eB)$ 
(blue line) shows some similarity with the CEP of the light quarks (red line) by 
moving to lower $\mu_B$. 
For the constant coupling $G_s^0$ scenario (magenta line) the CEP goes to 
higher values of $T$ but lower values of $\mu_B$.

Finally, some considerations concerning the deconfinement transition.
In the presence of a magnetic field the deconfinement transition is still a 
crossover, having an analytic behavior in opposition to a first-order phase 
transitions. The crossover transition thus allows for different definitions of 
the pseudo-critical temperature. 
In the right panel of Fig. \ref {fig:Fig3}, we present the $(T,\mu_B)$ values 
where $\Phi(T,\mu_B) = 0.5$, which is a possible way of defining a
pseudo-critical temperature for deconfinement, with $G_s=G_s^0$ (full lines) and 
$G_s(eB)$ (dashed lines) for two magnetic field strengths:
$eB=0.3$ GeV$^2$ (black lines) and $eB=0.6$ GeV$^2$ (red lines). 
We notice that the locations of the deconfinement transition is quite 
insensitive to the presence of an external magnetic field for both models. 
Furthermore, the analytic nature of the transition is preserved throughout the 
phase diagram.

\vspace{0.5cm}
{\bf Acknowledgments}

\vspace{0.25cm}
This work was supported by 'Fundação para a Ciência e Tecnologia', Portugal, 
under the project No. UID/FIS/04564/2016 and under the Grants No. 
SFRH/BPD/102273/2014 (P.C.), and under the project CENTRO-01-0145-FEDER-000014 
(MF) through CENTRO2020 program.
This work was partly supported by `NewCompstar', COST Action MP1304.


\end{document}